# Two-dimensional cuprate nanodetector with single photon sensitivity at T = 20 K

Rafael Luque Merino[1], Paul Seifert[1,2], José Durán Retamal[1,3], Roop Mech[1], Takashi Taniguchi[4], Kenji Watanabe[5], Kazuo Kadowaki[6], Robert H. Hadfield[7], Dmitri K. Efetov[1,8,9]

1. ICFO - Institut de Ciencies Fotoniques, The Barcelona Institute of Science and Technology, Castelldefels, Barcelona, 08860, Spain
2. Institute of Physics, Faculty of Electrical Engineering and Information Technology (EIT 2), Universität der Bundeswehr München, Neubiberg, 85577, Germany
3. Catalan Institute of Nanoscience and Nanotechnology (ICN2), BIST & CSIC, Barcelona 08193, Spain
4. International Center for Materials Nanoarchitectonics, National Institute for Materials Science, 1-1 Namiki, Tsukuba 305-0044, Japan
5. Research Center for Functional Materials, National Institute for Materials Science, 1-1 Namiki, Tsukuba 305-0044, Japan
6. University of Tsukuba, 1-1-1 Tennodai, Tsukuba-shi 305-8572, Japan
7. James Watt School of Engineering, University of Glasgow, Glasgow, UK
8. Fakultät für Physik, Ludwig-Maximilians-Universität, Schellingstrasse 4, 80799 München, Germany
9. Munich Center for Quantum Science and Technology (MCQST), 80799 München, Germany

*E-mail: dmitri.efetov@physik.lmu.de

---

Detecting light at the single-photon level is one of the pillars of emergent photonic technologies. This is realized through state-of-the-art superconducting detectors that offer efficient, broadband and fast response. However, the use of superconducting thin films with low $T_C$ limits their operation temperature below 4K. In this work, we demonstrate proof-of-concept nanodetectors based on exfoliated, two-dimensional cuprate superconductor $Bi_2Sr_2CaCu_2O_{8-\delta}$ (BSCCO) that exhibit single-photon sensitivity at telecom wavelength at a record temperature of T = 20K. These non-optimized devices exhibit a slow ($\sim ms$) reset time and a low detection efficiency ($\sim 10^{-4}$). We realize the elusive prospect of single-photon sensitivity on a high-$T_C$ nanodetector thanks to a novel approach, combining van der Waals fabrication techniques and a non-invasive nanopatterning based on light ion irradiation. This result paves the way for broader application of single-photon technologies, relaxing the cryogenic constraints for single-photon detection at telecom wavelength.

---

Superconducting nanowire single-photon detectors (SNSPD's) constitute an established technology for broadband, sensitive and fast detection of faint optical signals. [1-3]. These detectors, based on nano-patterned superconducting (SC) thin films, provide high efficiency [4], fast response ($\sim$ ns) [2, 3] and excellent timing resolution ($\sim$ ps) [5]. SNSPD's can be integrated in free-space and fiber-coupled architectures [1, 2], as well as planar photonic circuits [6, 7]. However, their operating temperature is limited to T < 4K because of low critical temperatures $T_C$ of the used thin films, such as NbN or WSi. Currently, the highest operating temperature for an SNSPD is 11K, demonstrated in a $MgB_2$ nanostrip [8]. Expanding single-photon detection to higher operating temperatures, longer wavelengths and faster response motivates the search for new SC material platforms [9-11]. In particular, van der Waals (vdW) heterostructures of 2D materials [12-15] and cuprate high-$T_C$ superconductors [9, 16-19] have received interest as candidates for the development of next-generation SNSPD's with increased capabilities.

Cuprate superconductors offer SC $T_c$ >100K and so are natural candidates to push SNSPD technology to higher temperatures [9]. A cuprate-based SNSPD could operate above liquid nitrogen temperatures (77 K) offering a critical technological advantage. However, fabrication of cuprate detectors has proven to be very challenging, as these degrade rapidly under ambient conditions. Despite extensive research efforts, most cuprate nanostructures exhibit flux-flow behaviour characterized by smooth current-voltage (*I-V*) characteristics that preclude SNSPD operation [16, 20]. Sharp and hysteretic transport characteristics, as those of conventional SNSPD thin films, were reported for high-quality YBCO nanowires along with the observation dark counts [18, 19]. To date, conventional fabrication approaches based on lift-off methods or heavy ion milling have not realized the prospect of a cuprate SNSPD.

In this manuscript we explore an alternative approach based on exfoliated, 2D cuprate $Bi_2Sr_2CaCu_2O_{8-\delta}$ (BSCCO). Novel fabrication methods allow us to harness the pristine superconducting properties of few-layer BSCCO [21, 22]. We report multiple nanostructures with sharp, hysteretic *I-V* characteristics combined with high $T_C$ > 70 - 80K. Optoelectronic measurements of the BSCCO nanostructures reveal single-photon sensitivity up to a record temperature of T = 20K.

**Device fabrication and transport characterization.**

Figure 1a summarizes the detector concept and fabrication. We fabricate van der Waals (vdW) heterostructures of few-layer BSCCO encapsulated with a top hBN inside an Ar-filled glovebox [23]. We use a bottom contact approach with thin metallic electrodes to bypass lift-off methods, which lead to oxidation of few-layer BSCCO [22]. In order to preserve superconductivity in BSCCO nanostructures, we use a Helium Focused Ion Beam (He-FIB) to pattern our BSCCO/hBN heterostructures [24], which can locally modify the doping level (and electronic properties) of few-layer BSCCO [25-27]. Direct patterning of insulating BSCCO areas can restrict the supercurrent to flow exclusively along channels of crystalline, non-irradiated BSCCO that preserves high $T_C$. Importantly, for low ion doses, the He-FIB irradiation does not etch through the encapsulating hBN layer, so keeping the BSCCO protected from the environment (see Fig. S4). This non-invasive patterning technique allows us to produce high-quality nanostructures based on 2D BSCCO flakes. This method also provides very high lateral resolution, allowing the writing of complex patterns with few nm resolution. The left inset in Fig. 1a shows an SEM image of a He-FIB defined nanostructure in our BSCCO/hBN stacks. This device is 250 nm wide and $2.7\ \mu m$ long. See Methods and sections S1 and S2 of the Supp. Info. for further details on the device fabrication.

We study the transport properties of the fabricated BSCCO nanostructures and compare them to pristine BSCCO flakes. In Fig. 1b we present temperature dependent measurements of the 2-terminal resistance *R*, normalized to its value at 120 K ($R_{120K}$) for three BSCCO/hBN nanopatterned detectors and an exfoliated, unpatterned BSCCO flake (reference sample). All the curves show clear superconducting transitions, where $R/R_{120K}$ drops to close to 0, with a residual resistance value stemming from the contact resistance in this 2-terminal device. The critical temperature of the BSCCO nanostructures $T_c$, defined from the mid-point of the superconducting transition, ranges between T = 60-80K. The lowered $T_c$ and broadened

SC transition of the patterned flakes likely derive from microscopic disorder and/or He-ion implantation [26, 27]. Still, the critical temperatures remain close to their nominal value and near cryogenic temperatures achievable with liquid nitrogen cooling.

Fig. 1c shows the *I-V* characteristics of the BSCCO nanostructures for different temperatures, and the inset shows the corresponding colour plot for the full temperature range. The *I-V* curve at *T* = 10 K features a sharp jump to the resistive state at a critical current $I_c = 280\ \mu A$. The combination of a high $T_c$ with a low critical current $I_c$ proves that the BSCCO nano-constriction confines the supercurrent into a very narrow channel, as the intrinsic $I_c$ of an unpatterned BSCCO flake is several orders of magnitude larger. The critical current density of the nano-constriction is $j_c$ ~ 7 x 10$^6$ A/cm$^2$, comparable to previous reports on high-quality nanowires based on YBCO thin films [18]. The vertical voltage jump at the resistive transition further reflects the high quality and low-disorder nature of our BSCCO nano-constrictions. As *T* increases, $I_c$ becomes smaller and the *I-V* characteristics are rounded by thermal noise. However remarkably, a sharply defined $I_c$ persists up to temperatures of *T* = 30 K.

Strikingly we observe a strong hysteresis in the *I-V* curves with respect to the sweeping direction of the bias current *I*$_b$, as shown in Fig. 1d at *T* = 20 K. As *I* is decreased below *I*$_c$, the superconducting state in the nanostructure is not recovered. Instead, the device switches back into its superconducting state at a much lower retrapping current *I*$_r$. The occurrence of the hysteresis is linked to the sharp *I-V* characteristics at *I*$_c$ and the high normal state resistance in the BSCCO nanostructure [18]. It also intricately depends on the thermal properties of the superconducting devices [19]. In our nano-structured BSCCO constrictions, the thermal conductance is reduced both in-plane, across the disordered irradiated BSCCO regions [28], and out-of-plane, as the 2D flake lies on the substrate without any lattice matching [29]. Figure S5 demonstrates the ubiquity of hysteretic behaviour in these nanostructures.

Such hysteretic behaviour is a key characteristic of state-of-the-art superconducting nanowire single photon detectors. The nucleation of a self-sustaining resistive region after photo-absorption induces Joule heating, precluding the recovery of the superconducting state when approaching I$_C$ from above. Superconducting nanowire detectors are kept well below T$_C$ and biased close to their critical current *I*$_b$ ~ *I*$_c$. Photo-induced "hotspots" are nucleated and grow due to Joule heating leading to a thermal runaway of the detector, which latches into the resistive state until the bias current is shunted [30]. Thus, a voltage drop takes place in the shunting circuit due to the resistive switching of the nanodetector. Remarkably, the observed hysteretic behaviour of the BSCCO nano-constrictions persists up to *T* = 30 K (see Fig. 1d) highlighting their potential as superconducting nanodetectors with single photon sensitivity and a high operating temperature.

**Photo-induced switching events in the BSCCO nanostructure.**

We now characterize the optoelectronic response of the BSCCO nano-constriction at a temperature of *T* = 20 K, where Fig. 2c shows the electronic read-out circuit and the SI Section 4 illustrates the optical schematics. The BSCCO nano-constriction is current-biased close to its switching current *I*$_b$ ~ *0.97 I*$_c$ and connected in parallel to a load resistor *R*$_{Load}$. The load resistor shunts the current once the nanostructure turns resistive, and enables free-running detection as the device self-resets after a time *τ*. We illuminate the biased BSCCO nanodetector using a CW laser at a telecom wavelength of $\lambda$ = 1550 nm, and monitor the voltage drop across $R_{Load}$ over time. We observe clear photo-induced clicks that appear only at elevated photon fluxes and are almost absent in dark conditions. Fig. 2a illustrates the appearance of these switching events under laser illumination.

Using an oscilloscope we measure the time-domain traces of the photo-induced clicks. We find that each click (Fig. 2b) features a fast rise time (few μs) and a slow decay time $\tau$ = 712 μs, obtained fitting the falling edge of the pulse. The full recovery time of the voltage signal lasts approximately 2ms. The timescales are consistent throughout all devices (see Fig. S7). The detector's $\tau$ is much slower than those of conventional SNSPD's ($\sim ns$) [2, 3]. The timescale is also slower than the intrinsic photoresponse time in cuprate superconductors ($\sim ps$) [31], aided by their high thermal conductivity κ [32].

We ascribe the observed timescale to the suppression of thermal transport in our BSCCO nanostructure, where thermal conductivity κ is reduced both in-plane and out-of-plane. The amorphous, insulating regions of irradiated BSCCO serve as a barrier for thermal relaxation, in addition to the surrounding areas of the flake, still in a superconducting state. In addition, out-of-plane thermal transport is also constrained by the vdW gap between the BSCCO flake and the substrate, as well as the lack of lattice matching. We believe the suppression of κ enables our nanostructures to exhibit sharp voltage jumps through an avalanche-like effect, instead of the smoother, flux-flow characteristics that are ubiquitous in cuprate nanostructures [16, 20]. It's worth noting that the detectors evolve towards flux-flow behaviour as the temperature is increased (see Fig. 1c). This is likely due to the enhancement of thermal transport and the softening of the gap for increasing temperatures. Lastly, our non-optimized current-shunting scheme may also lead to some bottleneck effect that limits the detector's relaxation time. Optimization of the device characteristics (both electronic and thermal) and the readout scheme could lead to enhanced performance in terms of detection speed.

We can identify the location from which the clicks originate by scanning the laser beam across the device. Indeed, the detector's photo-response (measured using the optical chopper) peaks around the expected position of the narrowest area of the nanostructure, as can be seen from the overlay of the photo-voltage map $V_{ph}$ with the scanning electron microscopy (SEM) image in Fig. 2c. Having identified the detector's active area, we now measure its switching rate using a Zurich UHLFI as an unreferenced, fast voltmeter. Fig. 2d shows the measured switching rate of the detector across the nano-constriction (indicated by the red dashed line in Fig. 2c). As the laser spot illuminates the "wings" of the nanostructure, the switching rate remains constant. The switching rate increases five-fold at the nanostructure's narrowest, most sensitive area. The high spatial sensitivity of the switching rate suggests that the clicks do not originate from a heating effect, rather from photo-induced Cooper pair breaking and a subsequent avalanche effect in the detector's active area. Overall, these observations establish that the observed clicks originate from photo-absorption in the active area of our nanostructure.

**Statistics of the switching events and single-photon sensitivity.**

We characterize the sensitivity of our BSCCO nanodetectors by studying the switching rate as a function of applied bias current $I_b$ and incident optical power $P$. All following measurements are performed at T = 20 K. In order to reliably discern switching events from the noise floor, we define a voltage threshold based on the detector's switching statistics under no illumination at each $I_b$. The following discussion is consistent for different choices of this threshold. The protocol to define the threshold is further detailed in section S6 of the SI.

First, we fix the optical power on the detector and vary the applied bias current $I_b$. For SNSPD-type detectors, the sensitivity and detection efficiency strongly depend on $I_b$, increasing as $I_b$ approaches $I_c$. These detectors reach single-photon sensitivity for $I_b \approx I_c$, at the cost of increased false counts due to thermal fluctuations and/or stray photons. Figure 3a shows the switching rate of the detector for increasing $I_b$ under P = 60 pW illumination and in dark conditions. We find that the switching rate of our BSCCO nanowire rapidly grows for $I_b \geq 0.9\ I_c$. False switching events also increase as the nanowire approaches its critical current. Notably, the onset of the dark switching rate takes place at a higher bias around $I_b \approx 0.95\ I_c$. This is consistent with a low-efficiency, current-biased SNSPD-type detector [1, 2]

A full saturation of the detector's switching rate, indicative of full internal quantum efficiency, was not observed. The growth of the switching rate slows down significantly for $I_b \geq 0.97\ I_c$, suggesting that a full saturation could be achieved. The non-saturating switching rate is common in nanoconstriction detectors where the device geometry leads to a position-dependent $I_C$. Thus, as $I_b$ is increased the effective active area of the nanodetector changes and the switching rate does not fully saturate [33, 34].

We analyze the statistics of the observed switching events at different optical powers for fixed temperature (20 K) and bias current $I_b$. At each point, we record switching events in a 5-minute window. Few-photon detection in SNSPD's is governed by Poissonian statistics, as the switching events are rare and uncorrelated. In Figure 3b, we illustrate the distribution of switching events at $I_b$ = 0.97 $I_c$ for increasing optical power. At low optical power (1 pW), very few clicks are observed and the switching events exhibit a highly skewed Poissonian-like distribution. As the optical power is increased, the switching rate of the detector increases and the distribution shifts towards higher mean values μ and higher variance $\sigma^2$. The distribution of the observed switching events is compatible with Poissonian statistics arising from few-photon clicks in the BSCCO nanodetector.

Lastly, we present the power dependence of the switching rate of our BSCCO superconducting nanodetectors. As the sensitivity of SNSPD-type detectors depends strongly on the applied bias current [34, 35], we study the sensitivity of our devices at 2 different bias conditions, highlighted by the vertical color segments in Fig. 3a. The low-bias condition (orange) corresponds to $I_b = 0.85\ I_C$ and the high-bias condition (blue) is $I_b = 0.97\ I_C$.

Figure 3c shows the power dependence of the switching rate in the high and low-bias conditions. The few-photon sensitivity of an SNSPD-type detector can be inferred from the power dependence of its switching rate. The probability to detect $k$ photons (switching event caused by simultaneous absorption of $k$ photons) is given by $P[k] \propto e^{-\lambda}\lambda^k/k!$, where λ is the mean (absorbed) photon number in the detector per unit time. For a single-photon detector the switching rate should increase linearly with incident power $P[k=1] \propto \lambda$. We observe a linear ($m = 0.95 \approx 1$) power dependence of the switching rate in the high-bias condition ($I_b$ ~ 0.97 $I_c$). When the bias current is lowered, the switching rate of the BSCCO detector is strongly reduced and exhibits a markedly different dependence. For low bias ($I_b$ ~ 0.85 $I_c$), the dependence falls closer to a power law of exponent $m = 2.93 \approx 3$, indicative of multi-photon (2 and 3-photon) events dominating the detector's response. Extended data supporting this observation is presented in the SI.

This behaviour is overall consistent with a single-photon sensitivity of the BSCCO nano-constrictions at $T$ = 20 K for telecom photons, where the observation of multi-photon sensitivity at lower bias current further supports the SNSPD-like characteristics of the BSCCO nanodetectors [34, 35]. As discussed before, far from the superconducting transition, the energy deposited by a single absorbed photon is not enough to trigger a switching event and multiple simultaneous photons need to be absorbed. Multi-photon sensitivity is rarely observed in commercial SNSPD's, as large fill factor designs make simultaneous absorption of multiple photons in the same area extremely unlikely. Instead, for detectors featuring constrictions, these events are more prominent and their photon sensitivity is strongly bias-dependent [34]. In our case, the CW nature of the illumination also promotes multi-photon contributions to the switching rate.

The detection efficiency can be roughly estimated using the incident power, number of switching events and the optical cross-section of the detector. Further details on the efficiency estimation are provided in Section S10 of the SI. In the high-bias condition where single-photon sensitivity is observed, we estimate a lower bound for the detector efficiency of $\eta \approx 3.3 \times 10^{-4}$. Several factors may contribute to this low figure-of-merit: i) due to the large superconducting gap of BSCCO, the energy of a single telecom photon would only break around a dozen Cooper pairs. The resulting quasiparticle generation process that leads to a resistive switching may therefore not be efficient [36]. ii) the effective (electronic) width of the active region may be smaller than estimated, as the microscopic details of the irradiation are unknown. iii) Another plausible scenario is the existence of a highly efficient, nanometric region inside the detector where photodetection takes place. Assuming an absorption coefficient of 17% [37] and full quantum efficiency, we can estimate the area of such a region to be $\sim 80\ nm^2$. This scenario is reminiscent of switching events around local microscopic defects and cannot be ruled out.

The microscopic details of the physical processes occurring after photo-absorption in the BSCCO nanostructures remain an open question. Even though our experiment does not probe the detection mechanism of this system, our results are consistent with an agnostic "hotspot" physical picture. In this picture, the photon energy is absorbed by the superconducting condensate and breaks Cooper pairs. The order parameter, or Cooper pair density, is locally suppressed in this "hotspot", precluding the supercurrent from flowing across the entire width of the nanostructure. Joule heating from the dissipation in this resistive region leads to an avalanche-like effect and superconductivity is lost in the nanostructure. This avalanche-like effect enables amplification of a microscopic perturbation (single-photon absorption) into a macroscopic voltage signal in SNSPD's.

The observation of strong hysteresis in the *I-V* characteristics of our nanodetectors supports the nucleation of a self-sustaining hotspot. We also observe latching of the nanodetector after photo-absorption in the absence of a shunting circuit, indicating that the hysteretic behaviour can be triggered optically. In literature, hotspots in cuprate thin films have been observed directly through thermal imaging [38] and indirectly through transport measurements [39, 40]. The large thermal conductivity κ of cuprates has limited cuprate nanostructures to the flux-flow regime, where an avalanche effect does not take place and ultimate sensitivity cannot be achieved. Suppression of the thermal transport in our 2D vdW nanostructures could explain the existence of a self-sustaining "hotspot". This behaviour is not expected to take place in Au-capped thin films studied in other reports [16, 18].

We believe that tailoring of thermal and electronic properties will be key for developing next-generation quantum sensors based on cuprate nanostructures. Achieving this while preserving the superconducting properties of the material may prove challenging. Other avenues to enhance detector performance include the design of optical gratings and nanoantennas to boost the absorption in the material. The integration in planar photonic architectures, highly compatible with 2D materials, could also boost the device's efficiency [6, 7, 22]. Lastly, the material properties of BSCCO also show promise for detection of single-photons at shorter wavelengths, as well as its application for photodetectors resilient to high magnetic fields [41].

In summary, we report single-photon sensitivity at $T$ = 20 K for telecom photons in nanostructures based on exfoliated 2D cuprate superconductor BSCCO. This proof-of-concept nanodetector sets the highest operation temperature reported to date for an SNSPD-type detector, establishing BSCCO-based vdW nanostructures as a promising alternative for next-generation quantum sensors.

## Acknowledgements

D.K.E. acknowledges support from the Ministry of Economy and Competitiveness of Spain through the "Severo Ochoa" program for Centres of Excellence in R&D (SE5-0522), Fundació Privada Cellex, Fundació Privada Mir-Puig, and the Generalitat de Catalunya through the CERCA program and funding from the European Research Council (ERC) under the European Union's Horizon 2020 research and innovation programme (grant agreement No. 852927). P.S. acknowledges support from the Alexander von Humboldt Foundation and the German Federal Ministry for Education and Research through the Feodor-Lynen program. K.W. and T.T. acknowledge support from the Elemental Strategy Initiative conducted by the MEXT, Japan (Grant Number JPMXP0112101001) and JSPS KAKENHI (Grant Numbers 19H05790, 20H00354, and 21H05233).

## Methods

### Device fabrication

We fabricate our samples through the dry-transfer technique inside an Ar-filled glovebox (H2O, O2 < 0.8 ppm). Few-layer flakes of optimally doped BSCCO (10 - 20 nm) are exfoliated using PDMS, then transferred onto pre-patterned metallic electrodes (Ti/Au, 2/20 nm). BSCCO flakes are then encapsulated using pre-selected thin hBN flakes (∼ 20 nm). We probe the 2-terminal resistance of our devices using a probe-station inside the glovebox. The entire process is performed in less than 20 minutes to avoid degradation of few-layer BSCCO flakes. The entire vdW assembly process is done at room temperature.

The chip is then pasted onto a stub for He-FIB using silver paste and the contacts of the device are bonded onto the metallic stub, so the entire device remains grounded during patterning. Outside the glovebox, the devices are always in vacuum using a commercial vacuum container. Therefore, the device is only exposed momentarily, for example during bonding.

In the He-FIB we work with 30 KeV acceleration voltage for the Helium beam, and typical emission currents between 2-5 pA. For each sample, we calibrate the optimal He ion dose in order to locally turn BSCCO insulating. This process is discussed in Section 2 of the SI, as well as in [22]. Typical ion doses range between 20-90 pC/$\mu m^2$. Crucially, for low ion doses, the Helium ion beam does not etch through the encapsulating hBN layer so the BSCCO flakes remain protected. We fabricate simple nanowires as well as complex structures (see Section 1 of the SI). Lastly, we remove the sample from the metallic stub and bond the chip to a pre-grounded PCB. The PCB is then loaded into our optical cryostat.

Our devices are typically kept in the cryostat for 2-week measurement runs. Careful thermal cycling can preserve the device's SC properties. Prolonged exposure to air leads to oxidation due to oxygen out-diffusion and water creep through the device-substrate interface.

**Transport & optoelectronic setup**

We characterize the optoelectronic properties of the vdW BSCCO nanowires using an Attodry800 optical cryostat with a base temperature of 8 K. For transport characterization, we perform both AC Lock-in measurements and DC measurements. For the AC Lock-In measurements, we use a 10 nA AC excitation current at a reference frequency of 126 Hz. An SRS860x Lock-In amplifier is used both for sourcing and sensing. DC measurements are performed using a combination of DC sources (Yokogawa GS200, Keithley 2450) and a multimeter (Keithley 2700). It is worth noting that for device C, the normal state resistance is very large (~ 100 KΩ), making a voltage bias circuit favourable over a current bias circuit.

Laser excitation is provided by a telecom CW diode laser (Thorlabs SFL1550P), driven by a DC current source (Keithley 2400). The laser beam is first attenuated using neutral density filters. Then, it is aligned and focused onto the sample using a combination of room-temperature galvanic mirrors and a cryogenic 3-axis piezo stage for the sample. At optimal focus, the laser has a beam width of $w_0 \sim$ 2 μm. The detector is current-biased, connected in parallel to a load resistor. The voltage drop at the load resistor generated by the detector's resistive switching is recorded using a Zurich UHFLI (600 MHz bandwidth, 1.8 GSa/s sampling rate). A low-noise amplifier is used to pre-filter the voltage spikes. The Zurich UHLFI records switching events for 5 minutes at each optical power and bias current setting. Time - domain traces of the switching events are recorded using the oscilloscope function of the Zurich UHFLI with a sampling rate of 175 MHz. Alternatively, we study the photovoltage signal of the device using an optical chopper and a Lock-In amplifier (SRS860x) referenced at the chopper frequency. Optimizing the optical coupling to the detector becomes easier in this "slower" configuration (see Fig. 3c). A schematic of the optoelectronic setup for single-photon detection is shown in Fig. S6.

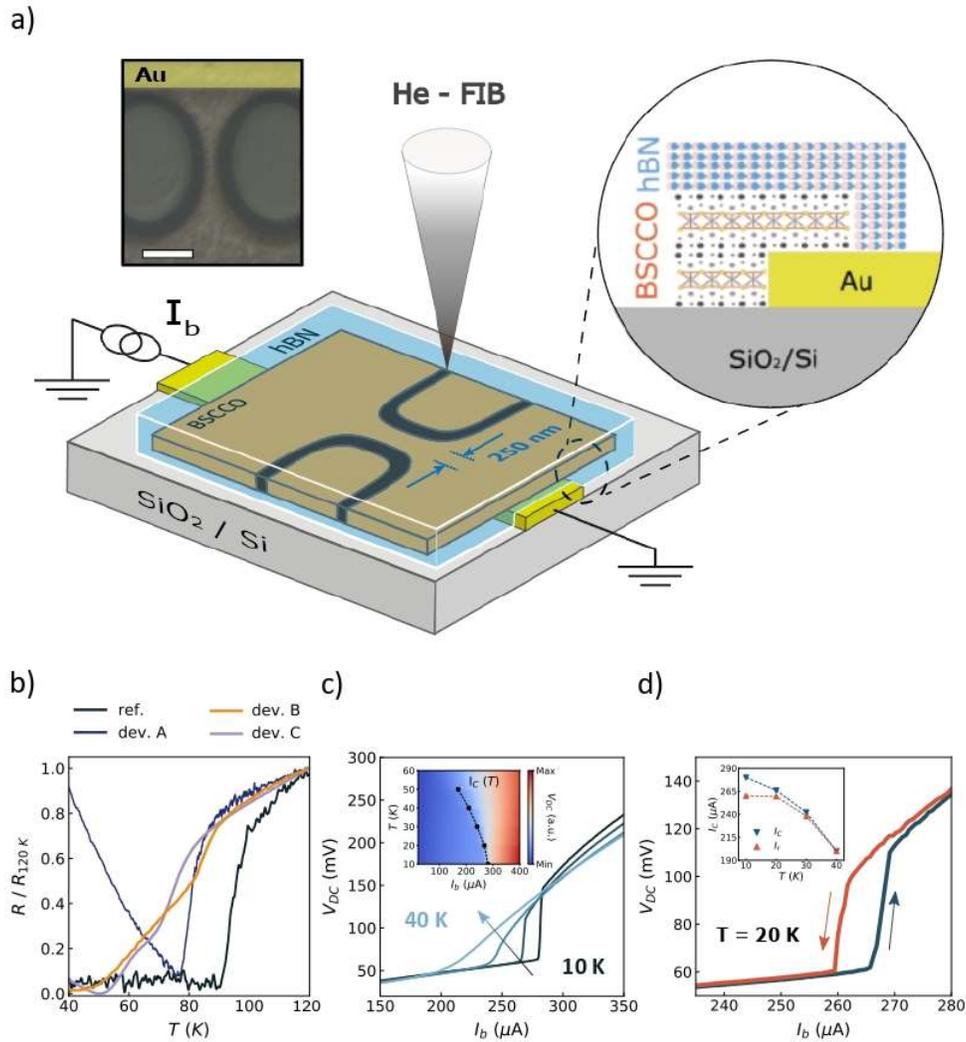

*Figure 1. Two-dimensional superconducting cuprate nanowires for photodetection. a Illustration of the detector concept. A high-$T_C$ superconducting BSCCO flake is encapsulated with hBN and then nanopatterned using a focused He-ion beam. The resulting nanostructure is current-biased ($I_b$) in a 2-terminal configuration. Left inset shows an SEM image of the nanowire in device A, scale bar has 1 μm length. Right inset depicts a schematic cross-section of the vdW heterostructure. b 2-terminal resistance vs. temperature of the BSCCO nanostructures. All curves have been normalized to the resistance value at 120 K, before the superconducting transition. Black curve shows the 4-terminal resistance for a reference, unpatterned BSCCO flake. For all nanostructures, $T_C$ ranges between 60 – 80 K. c I-V characteristics of device A for increasing temperature. The sharp voltage jump at 10 K evolves towards a smooth transition at 40 K. Inset depicts the superconducting phase diagram for this device. Overlayed plot indicates the extracted $I_C$ (T) dependence. Beyond 50 K, the smooth I-V characteristics preclude a reliable extraction of $I_C$. d I-V characteristic for device A at T = 20 K, exhibiting a sharp voltage jump together with a well-defined hysteresis loop. Vertical offset between red and blue traces is due to Joule heating effects in the DC measurement. Inset shows $I_C$ (T) and $I_r$ (T). Hysteresis loop remains finite up to T = 30 K.*

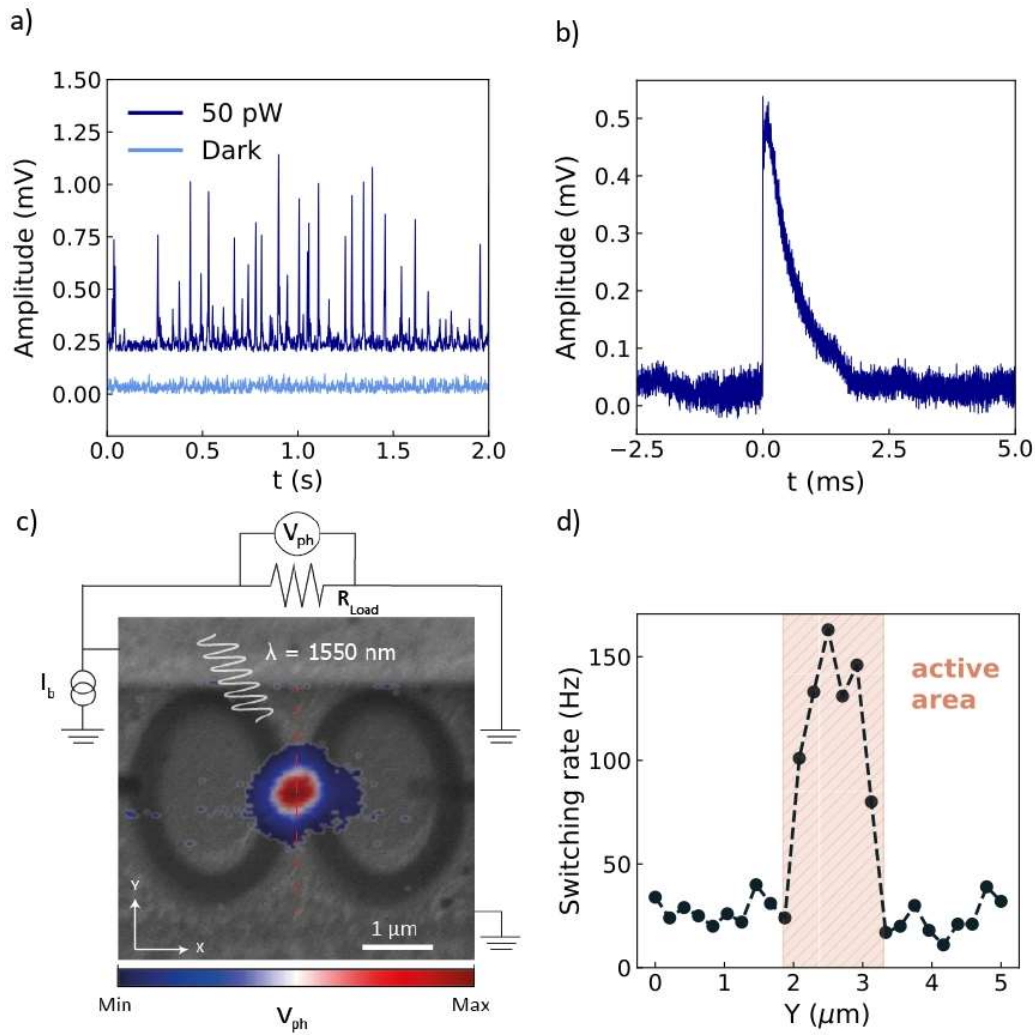

***Figure 2. Photo-induced voltage signal in Bi2212 nanowires. a** Photo-induced voltage drop at the load resistor over time for 50 pW optical power (black) and under no illumination (grey). Black trace is offset by 0.2 mV for clarity. Traces recorded at T = 10 K, $I_b = 0.98\ I_C$. **b** Oscilloscope trace of the voltage trace generated by a switching event. The obtained decay time is $\tau = 712\ \mu s$. Switching rate of the detector along the red dashed line in **(a)**. Optoelectronic measurement schematic and spatial photovoltage map, overlayed with SEM image of the nanowire. The nanowire is biased close to its critical current $I_b \sim I_C$. **d** Switching rate of the detector along the red dashed line in **(c)**.*

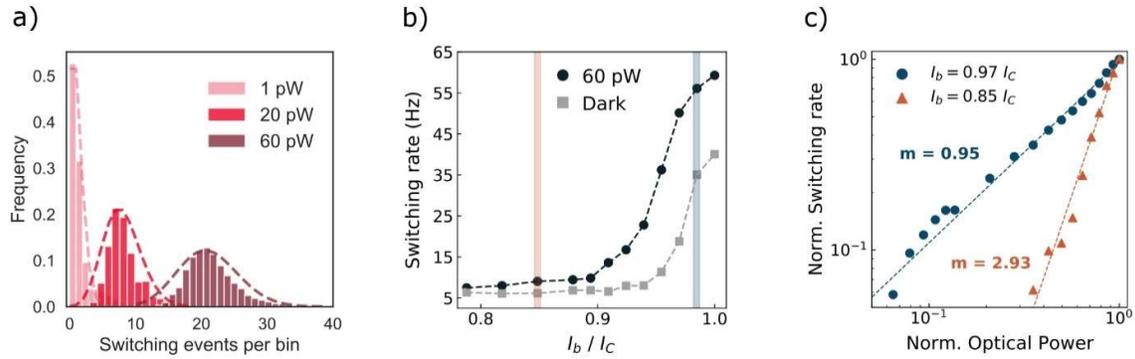

*Figure 3. Single photon sensitivity at T = 20 K. a* Switching rate and dark count rate of the nanodetector at 20 K for increasing bias current. Vertical segments indicate the bias points examined in *(c)*. *b* Distribution of the photo-induced switching events for increasing optical power at 20 K, $I_b = 0.97\ I_C$. The switching event distribution is consistent with Poissonian statistics of few-photon events. Dashed lines illustrate the best-fitting Poissonian distribution. *c* Normalized switching rate versus normalized optical power for two different bias conditions. Each magnitude is normalized to its maximum value. The switching statistics for $I_b = 0.97\ I_C$ follow a linear dependence, indicating single-photon sensitivity of the nanodetector. For $I_b = 0.85\ I_C$, the switching rate follows a power law of exponent m ≈ 3, indicating multi-photon sensitivity of Device A at lower bias currents. Switching events recorded at 20 K. The maximum switching rate for $I_b = 0.97\ I_C$ and $I_b = 0.85\ I_C$ are 450 Hz and 38 Hz, respectively. Maximum optical power is 100 pW.

# Supplementary Information

## Table of contents

## Supplementary Section 1: Device fabrication

In Figure S1, we present optical images for the 3 reported devices at different stages of our fabrication process for air-sensitive, two-dimensional BSCCO nanostructures. Namely, the pictures depict the stacking of the BSCCO flakes onto prepatterned gold electrodes (left column), the finalized stack with hBN encapsulation (central column) and the He-FIB defined nanostructures written in these vdW stacks (right column). The first row (S1a-S1c) corresponds to Device A, the second row (S1d-S1f) corresponds to Device B and the third row (S1g-S1i) corresponds to Device C. As seen in Fig. S1f, multiple 2-terminal channels can be defined in a single BSCCO/hBN heterostructure.

Devices A and C use a digitated electrode design. The digitation of the metallic electrodes is meant to increase the contact area between the BSCCO and the edges of the metallic pads. As the outer layers in the BSCCO unit cell are insulating, a purely planar bottom contact does not yield optimal electrical contact. The digitated design serves as a compromise between maximum contact area plus contact area at the electrode edges, where we believe the electrons have a higher probability of accessing the copper oxide planes in BSCCO.

We determine the thickness of the BSCCO flakes through Atomic Force Microscopy (AFM). The AFM scans are performed after completion of the experiments. Figure S2 depicts the AFM scans for devices A and B. The thickness of the BSCCO flake in Device A is $d = 16.5\ nm$ (5.5 u.c.). The thickness of the BSCCO flake for Device B $d = 15\ nm$ (5 u.c.), respectively. Device B was damaged after the measurement cycle, impeding a reliable estimation of the flake thickness. From the optical contrast of the BSCCO flake, we estimate its thickness to be around $d = 21\ nm$ (7 u.c.). For ultrathin layers like bilayers ($3\ nm$) and trilayers ($4.5\ nm$), we found that the contact resistance increased dramatically, with typical values between $0.1 - 10$ MΩ. Some of these devices did show signatures of superconductivity, but such large contact resistance prohibits the use of these flakes to build nanodetectors.

Naturally, such ultrathin layers are even more prone to oxidation during the fabrication process.

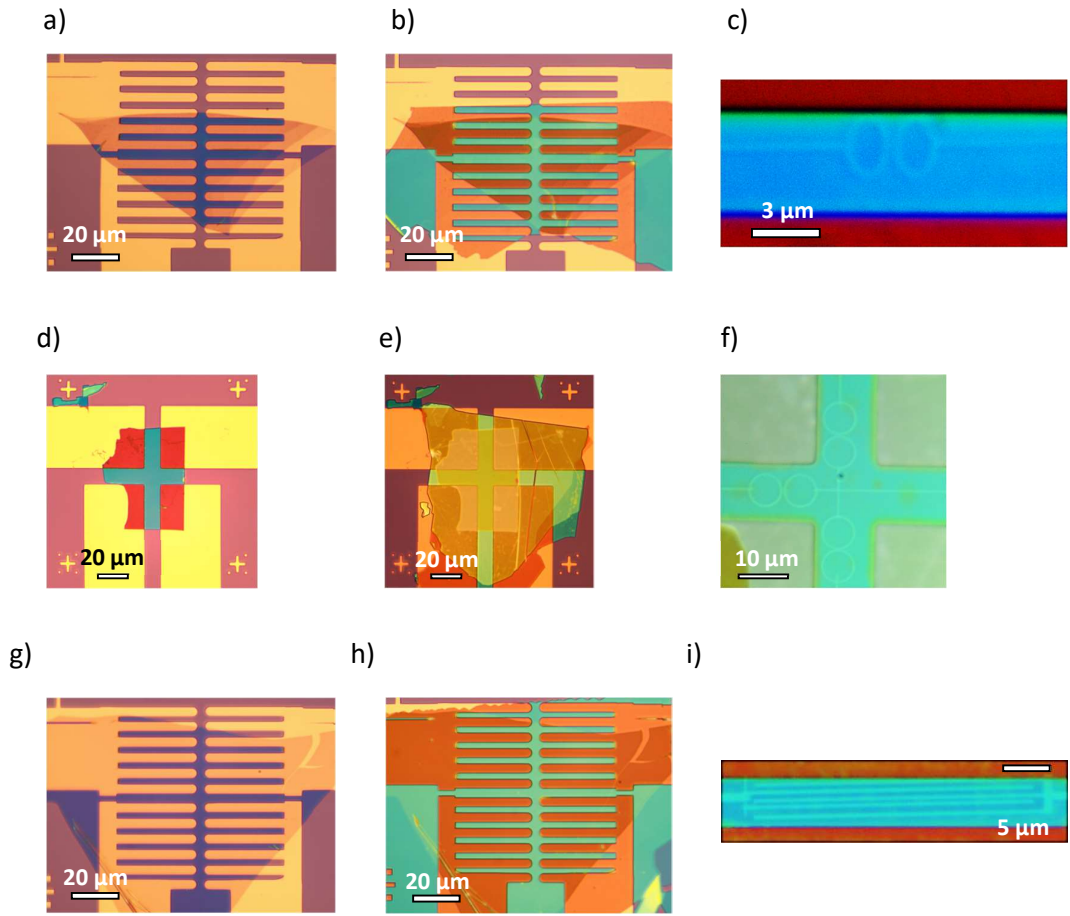

**S1. Optical images of the reported BSCCO/hBN heterostructures. (a-c)** show Device A. The nanowire in **(c)** has width w = 250 nm and length l = 2.7 µm. **(d-f)** show Device B. As shown in **(f)**, multiple 2-terminal devices can be written on a single vdW heterostructure. The nanowire on the left in **(f)** (reported in this study) has width w = 300 nm and length l = 4 µm. Figures S1 **(g-i)** show Device C. The meandering nanowire shown in **(i)** has width w = 500 nm and total length l = 120 µm

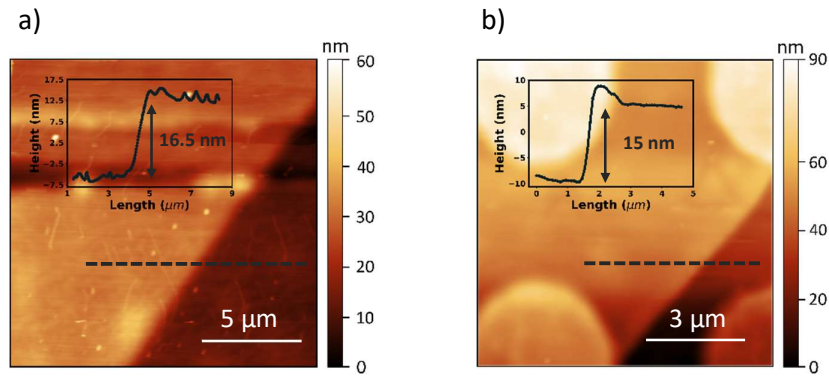

**S2. Topographic images of the few-layer BSCCO flakes. a** AFM image of Device A. Inset: Height profile along the black dashed line. The thickness of the BSCCO flake is d = 16.5 nm, which corresponds to 5.5 unit cells. **b** AFM image of Device B. Inset: Height profile along the black dashed line. The thickness of the BSCCO flake is d = 15 nm, which corresponds to 5 unit cells.

## Supplementary Section 2: He-FIB nanopatterning

Here, we describe the protocol for calibration of the ion dose in the He-FIB nano-patterning process, as well as the topographic profile of the irradiated areas.

The irradiation of BSCCO with a tightly focused beam of Helium ions leads to a local modulation of its electronic properties, through a local change in the doping level [1]. However, in order to produce high-quality superconducting BSCCO devices, the ion dose must be minimized in order to avoid physical damage. The sample remains grounded during the patterning process, leading to a bright contrast in the secondary electron imaging provided by the He-FIB tool. In order to optimize the ion dose, we pattern triangular shapes with increasing ion dose onto our BSCCO/hBN heterostructures. Initially, the interior of the triangles remains bright as the irradiated BSCCO areas are not resistive. As the ion dose increases, the interior of the triangles becomes dark. This is due to the accumulation of positive charge inside the triangle that can't be compensated through the ground; as now the irradiated perimeter of the triangle is an electrical insulator. This process is illustrated in Figure S3 for low and high ion doses in Device A. More details can be found in Ref. [2].

At this point, we verify the non-etching character of this patterning technique. We study the height profile of our irradiation-defined superconducting nanostructures in BSCCO/hBN stacks. Figure S4 shows an AFM scan of the studied nanostructure in Device A. For an ion dose of $100\ pC/\mu m^2$, the surface exhibits a height change of $4 - 5$ nm. The thickness of our encapsulating hBN layer is much larger (20-25 nm) than the height change induced by the irradiation. Then, we can conclude that our nanopatterning process is indeed non-invasive.

In previous reports of He-FIB irradiation of cuprates, authors observed an amorphization of the crystal structure of BSCCO in the irradiated areas [3]. It is likely that such amorphization also takes place in our devices. However, the transport characteristics of our our devices (Figure 1, Figure S5) point towards crystalline, high quality (super)conducting nanostructures [4] which are surrounded by insulating barriers of amorphous BSCCO. The effects of this amorphous surrounding matrix on the thermal properties of these devices is an important open question for development of High-Tc devices with light-ion irradiation.

Heavy-ion focused ion beams have already been explored as a way to define cuprate nanostructures [5]. However, the large mass of the used ions (Ga+, Ne+,…) set a limit on both the quality of the superconducting properties and the nanostructure dimensions. The lack of pristine superconducting properties and/or nanoscale dimensions limited these devices to the flux-flow regime.

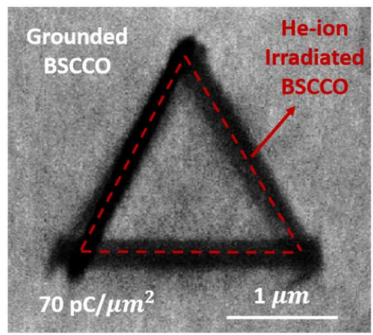 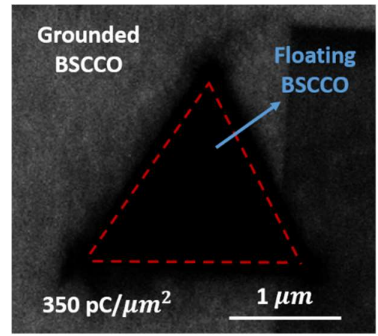

**S3. Secondary electron image of He-FIB irradiated triangular patterns on Device A. a** Ion dose of 70 pC/$\mu m^2$ **b** Ion dose of 350 pC/$\mu m^2$.

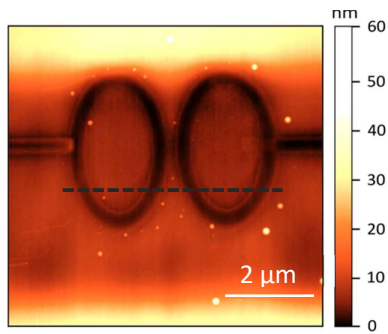 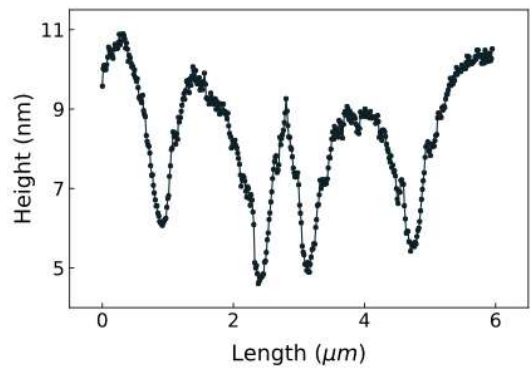

**S4. Non-etching nanopatterning of BSCCO flakes. a** AFM image of He-FIB defined pattern on Device A. The higher (brighter) regions correspond to the metallic electrodes. **b** Height profile across the dashed line in **(a)**. The irradiated region exhibits a change in height of approximately 4 - 5 nm.

## Supplementary Section 3: Transport characterization BSCCO nanostructures

In this section we discuss the general transport characteristics of the BSCCO nanodetectors. After the vdW assembly in the glovebox, we check the 2-terminal resistance along all contact combinations. Non-oxidized sample with good contact exhibit 2-terminal resistance below 10 KΩ. After nanopatterning the 2-terminal resistance showed an increase for all devices, due to the confinement of the current and some degradation during the minimal air exposure of the devices. From Wiedemann-Franz law, an increase in the device resistance comes with a decrease in its thermal conductivity which supports our scenario of self-heating "hotspots" appearing in the BSCCO nanostructures.

The critical temperature $T_C$ (after nanopatterning), defined from the mid-point of the superconducting transition, remains above T = 70 K for all 3 devices.

The extracted critical temperatures are:

$T_C$ (Reference device) = 94 K  $T_C$ (Device A) = 81 K

$T_C$ (Device B) = 77.5 K  $T_C$ (Device C) = 73K

In Figure S5 we report the hysteretic *I-V* characteristics of the 3 devices. A parallel load resistor is not used in these measurements. The nonlinear *I-V* characteristics at low applied DC current are due to the suppression of the Schottky-like barrier at the vertical BSCCO/Au contact. All 3 devices exhibit a sharp voltage jump and hysteretic behaviour, signalling high quality material properties.

Note that the observation of critical currents between 200 µA and 500 µA (while $T_C$ remains high) directly implies the confinement of the supercurrent to a narrow channel, i.e. the patterned nanostructure. If the supercurrent were not confined to the channel, $I_C$ would be orders of magnitude higher, easily reaching $\sim 1\ A$, as BSCCO exhibits a very high critical current density $J_C$.

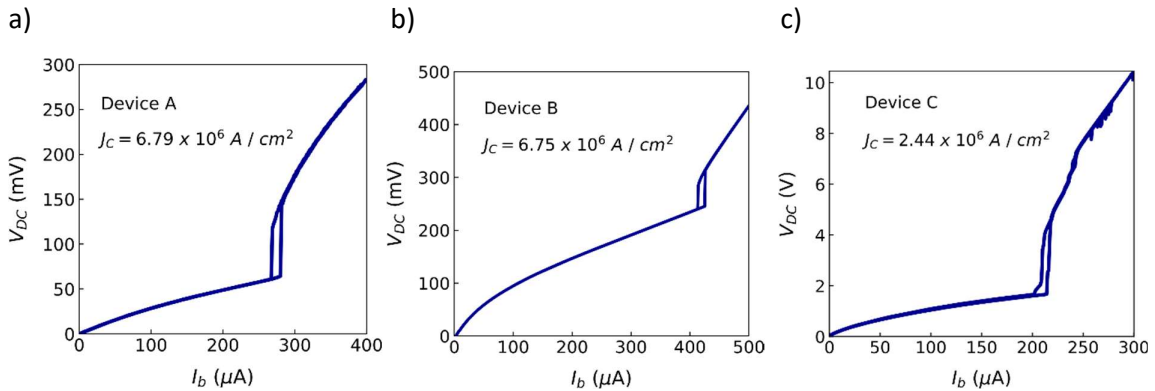

**S5. Hysteretic I-V characteristics for BSCCO nanostructures. a** -IV characteristic for Device A at T = 10 K. **b** I-V characteristic for Device B at 15 K. **c** I-V characteristic for Device C at 10.8 K. Multiple voltage jumps correspond to the switching of different regions within the meander. The normal state resistance is also much higher.

## Supplementary Section 4: Optoelectronic setup

Figure S6 shows a schematic of the optoelectronic setup. A continuous wave (CW) single-frequency laser source (Thorlabs SFL1550P) is driven using a sourcemeter (Keithley 2400) to provide laser excitation of the detector. Neutral density (ND) filters are used to attenuate the optical excitation. The laser excitation is kept at a linear polarization. The laser beam may be modulated using an optical chopper for AC Lock-In measurements of the detector's response as done in Ref [2]. The beam is scanned in the XY plane using galvanic mirrors, before being focused onto the sample by a low-temperature objective (Olympus LCPLN50XIR, NA = 0.65). In order to locate the active area of the detector and optimize the free-space coupling, we move the sample using a 3-axis piezo positioner stack. The sample is located in an Attodry 800 optical cryostat with a base temperature of 8 K. The superconducting nanowire is current-biased using a sourcemeter (Keithley 2450) in series with a large bias resistor. A load resistor is connected in parallel to the device in order to shunt the bias current after resistive switching of the nanostructure. We monitor the voltage drop over the load resistor using a Zurich UHLFI, which is used both as an unreferenced, fast voltmeter and as an oscilloscope probe. We record long traces of the voltage drops using it as a voltmeter, while the oscilloscope probe is used to inspect the time-domain traces of the switching events. A Lock-In amplifier (SRS860x) is used for measurements of the photovoltage at the modulation frequency of the optical chopper $f_0$.

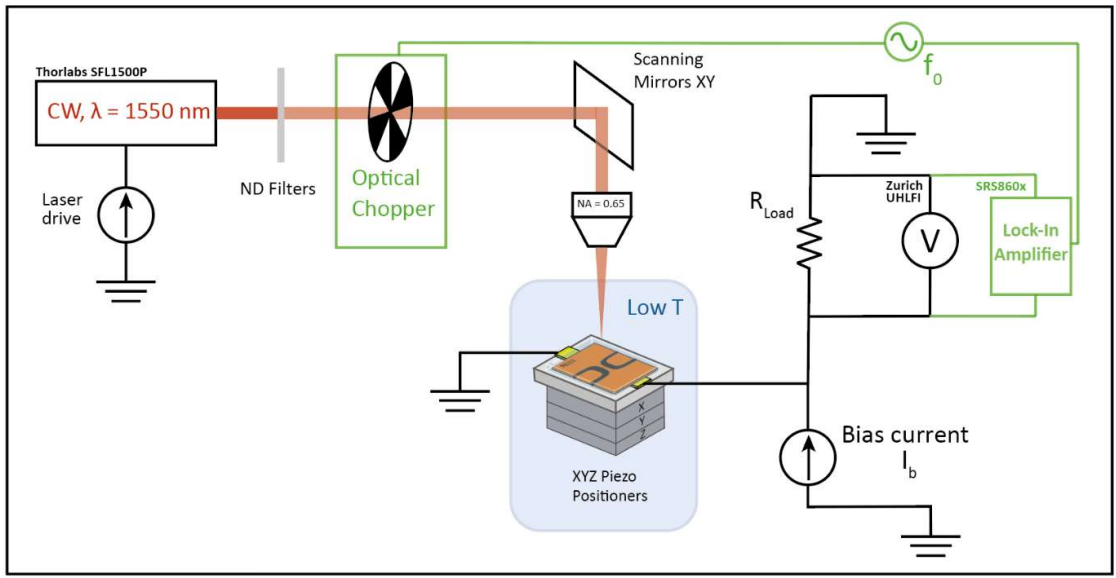

***S6. Schematic of the optoelectronic setup.*** *Elements highlighted in green are only used for AC Lock-In measurements of the photoresponse.*

## Supplementary Section 5: Traces of photo-induced switching events

Figure S7 shows time-domain traces of the voltage drops at the load resistor caused by the resistive switching of the BSCCO nanodetector. For the 3 devices, the decay time of the voltage pulses is of the order of 1 millisecond. Differences in the noise levels of the traces are due to different signal-to-noise ratios for the signal of each device, as well as differences in the configuration of the oscilloscope probe. For the 3 devices, the typical amplitudes of the measured voltage drop range between 0.1 – 1 mV (before amplification). It is worth noting that for Device C a voltage bias circuit is used, eliminating the need for a parallel load resistor, but also modifying its mode of operation and photodetection statistics.

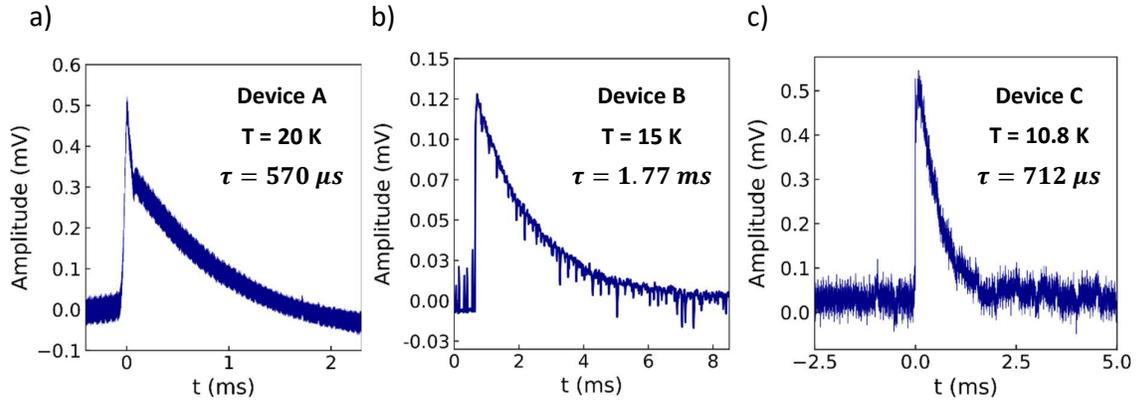

***S7. Click traces recorded with the oscilloscope probe for each device. a** Switching event voltage trace for Device A at T = 20 K. **b** Switching event voltage trace for Device B at T = 15 K. **c** Switching event voltage trace for Device A at T = 10.8 K. The decay time for all devices ranges between 0.5-2 ms.*

## Supplementary Section 6: Protocol for counting switching events

In order to study the statistics of these switching events, we record the voltage drop over the load resistor in 5 minute windows. To have a robust determination of the trigger level (threshold voltage for a switching event), we use the data from "dark" switching events (plotted in Figure 3b in the main text) as a function of bias current.

Our trigger estimation goes as follows: For each $I_b$ value, we record the voltage drop over the load resistor (under no illumination, optical window blocked) in 5 minute windows. Then, we construct a histogram of the voltage amplitudes in this dataset (both "dark" switching events and points in the noise floor). The Gaussian distribution of the amplitudes is fitted to extract its mean µ and standard deviation σ.

We define the trigger level at each $I_b$ value as $trigger(I_b) = \mu(I_b) + N\sigma(I_b)$, where N acts as a tuning knob to modify the trigger level. In summary, the trigger level is defined at each bias current $I_B$ from a percentile of the larger voltage values measured under no illumination. In Figure S8, we illustrate the trigger selection, as well as showing how the single-photon statistics reported in Figure 3c are robust to the voltage threshold. Figures in the main text make use of a trigger level where N = 3.5.

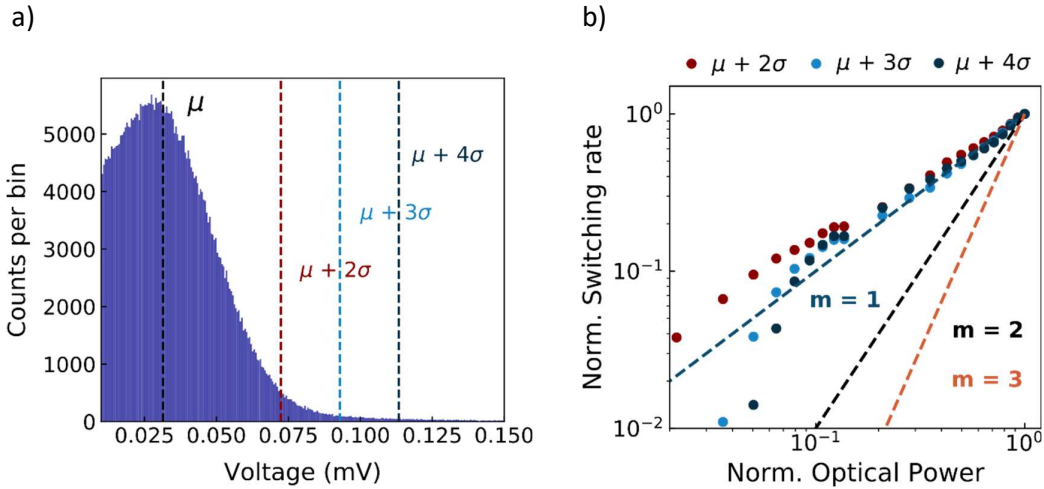

**S8. Determination of the trigger for switching events. a** Histogram of the voltage amplitudes without laser illumination at $I_b = 0.97\ I_C$. Vertical lines indicate µ and the trigger levels illustrated in **(b)**. **b** Power dependence of the switching rate at $I_b = 0.97\ I_C$ at different trigger levels. Beyond the dark count rate, the dependence remains linear Importantly, the linearity is robust versus the trigger choice.

## Supplementary Section 7: Single photon sensitivity at T = 20 K for Device A

Figure S8 summarizes the performance of Device A. In S8a, the device image is shown, along with a zoom-in of the nanostructure that exhibited single-photon sensitivity. Scale bars for these images can be found in Fig S1. Figure S8b depicts the statistical distribution of the switching events in the high-bias condition ($I_b = 0.97\,I_C$). Dashed lines illustrate the Poissonian-like statistics from these events. The inset depicts a trace of one of these switching events. Figure S8c illustrates the single-photon and multi-photon sensitivity of Device A under different bias conditions. In the inset, we use a linear-linear scale to emphasize the difference in the optical power dependences in the 2 bias conditions. Figure S8d shows a power-law plot of the optical power dependence of the switching rate at 4 bias conditions. For $I_b = 0.97\,I_C$ and $I_b = 0.96\,I_C$, the switching rate evolves linearly as expected for the single-photon sensitivity. However, both for $I_b = 0.90\,I_C$ and $I_b = 0.85\,I_C$, the power-dependence of the switching rate is best fit with a power law of exponent $m > 1$. Dashed lines indicate the best power-law fit. The exponents extracted from the fit are:

$$m\,(I_b = 0.85\,I_C) = 2.93 \pm 0.14 \qquad m\,(I_b = 0.90\,I_C) = 2.21 \pm 0.21$$
$$m\,(I_b = 0.96\,I_C) = 0.94 \pm 0.04 \qquad m\,(I_b = 0.97\,I_C) = 0.95 \pm 0.03$$

All data displayed in Figure S8 was obtained at a temperature of T = 20 K.

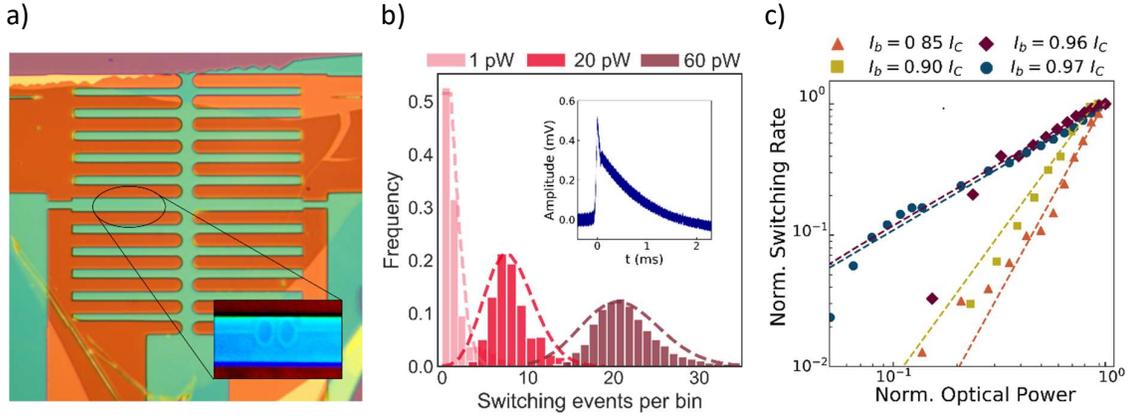

**S8. Single photon sensitivity at T = 20 K for Device A. a** Device image. Inset shows zoom-in of the studied nanostructure. Scale bars for both images are shown in Fig. S1. **b** Statistical distribution of the switching events in the high-bias condition. Inset shows a trace of a switching event. **c** Extended data of single and multi-photon sensitivity at different bias for Device A. Dashed lines correspond to the best power-law fits. Exponents indicated in the text.

## Supplementary Section 8: Single photon sensitivity at T = 15 K for Device B

In Figure S9, we present evidence for single-photon sensitivity for another device at a temperature of T = 15 K. The bias condition used in this device was $I_b = 0.98\, I_C$. Figure S9a shows Device B, as well as a zoom-in of the studied nano-constriction. Scale bars and dimensions of the nanostructure are shown in Figure S1. The statistical distribution of the switching events is shown in Fig. S9b, along with a time-domain trace of a switching event. As for Device A, the switching event statistics are consistent with the Poissonian character of few-photon detection. In Figure S9c, the optical power dependence of the switching rate is shown for $I_b = 0.98\, I_C$ at T = 15 K. Compared to Device A, we observe a wider background of "dark" switching rate around 20 Hz. For increasing optical power, the switching rate grows in a linear fashion. For comparison, the red dashed line illustrates cubic growth of the switching rate. Clearly, the switching rate statistics fit best with a linear dependence, indicative of single-photon sensitivity. Therefore, we conclude that Device B also shows evidence of single-photon sensitivity at T = 15 K very close to its critical current.

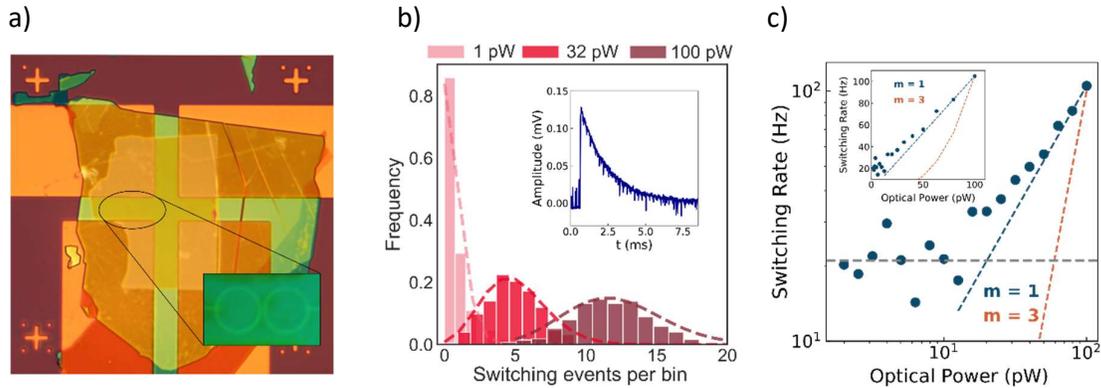

***S9. Single-photon sensitivity at T = 15 K for Device B. a*** *Device image. Inset shows zoom-in of the studied nanostructure. Scale bars for both images are shown in Fig. S1.* ***b*** *Statistical distribution of the switching events at $I_b = 0.98\, I_C$. Inset shows a trace of a switching event.* ***c*** *Optical power dependence of the switching rate of Device B at $I_b = 0.98\, I_C$ and T = 15 K. Inset depicts the power dependence in a linear-linear scale.*

## Supplementary Section 9: Photodetection statistics at T = 10.8 K for Device C

Device C features a meandering nanowire, resembling the high-fill factor structures used for commercial SNSPD's. The meandering nanowire has width $w = 500\ nm$ and total length $l = 120\ \mu m$. The transport characteristics are comparable to Device A and Device B, featuring a high critical temperature $T_C$ = 73 K and *I-V* characteristics with a sharp voltage jump and pronounced hysteresis. The normal state resistance of the device is much larger for Device C and the IV characteristic exhibits multiple voltage jumps, each one of them with its own hysteresis (Fig. S5c). The increased resistance follows directly from the larger nanostructure in Device C. The multiple voltage jumps likely stem from inhomogeneities throughout the nanostructure, defining several consecutive nanoconstrictions that switch at different critical currents.

Due to the high resistance of this device, it was not possible to current bias of the detector. The main (first) hysteresis loop in its *I-V* characteristic was not possible, as the hysteresis loop did not close for any choice of load resistor. Therefore, the detector latched after every switching event. We applied a voltage bias instead, without any load resistor in parallel to the device. In this case, it's the source itself that shunts the detector after switching. In this voltage bias configuration, switching events were clearly measured (with a better signal-to-noise ratio). However, the statistics of photodetection in the voltage-biased BSCCO meander deviated from the behaviour of devices A & B.

At this point, the nature of this response is not well understood. It is likely a combined effect from the different bias scheme, along with the complex meandering nanostructure where multiple "hotspots" could appear. Correlations between switching events could appear due to the appearance of simultaneous/consecutive hotspots within the nanostructure. We believe these dynamics may wash out the statistics from single-photon detection in this device. The data presented in Figure S10 was measured at T = 10.8 K and at 99% of the critical voltage $V_C$.

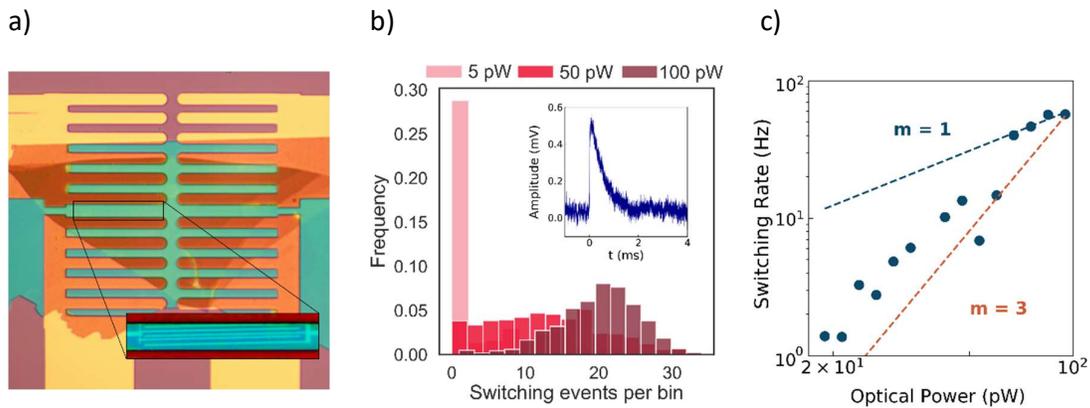

*S10. Photodetection statistics for Device C at T = 10.8 K.* **a** Device image. Inset shows zoom-in of the meandering nanowire. Scale bars for both images are shown in Fig. S1. **b** Statistical distribution of the switching events at $V = 0.98\ V_C$. The distribution of the switching events evolves differently from a Poissonian distribution. Correlations between small resistive areas within the meander could explain the non-Poissonian distribution of the switching events. Inset shows a trace of a switching event. **c** Optical power dependence of the switching rate of Device C at $V = 0.98\ V_C$ and T = 10.8 K. For high optical powers, we observe saturation and even decrease of the switching rate of Device C.

## Supplementary Section 10: Estimation of the detector's efficiency

In our experiment, several parameters required for a detailed estimation of the detector's efficiency remain undetermined. These include the exact electronic width of the irradiated nanostructures and the effects of the hBN capping layer, the substrate and the ion implantation. We provide a rough estimation of the detection efficiency considering the observed switching rate in the high-bias condition, the optical power entering the cryostat's window, the overlap between the device and the laser spot at this position and the absorption coefficient for few-layer BSCCO at 1550 nm.

For Device A, at T = 20 K, $I_b = 0.97\, I_C$ the switching rate is 445 Hz for 100 pW incident optical power. For a photon energy of 0.8 eV ($\lambda = 1550\ nm$), approximately $8\ x\ 10^8$ photons per second are incident on the detector. As the beam width at optimal focus (2 µm) is much larger than our nanowire's dimensions (Figure 2c in the main text), we calculate the coupling efficiency between our laser spot and the nanowire's active area. The active area of the nanowire is approximated as a rectangle of width $W$ = 250 $nm$ and length L = 1 $\mu m$. For the laser spot, we take a Gaussian beam with beam waist $\sigma = 2\ x 10^{-6}\ m$ at optimal focus. We compute the convolution between our laser spot and the detector's active area:

$$\frac{1}{2\pi\sigma^2} \int_{L/2}^{-L/2} \int_{W/2}^{-W/2} e^{-\frac{(x^2+y^2)}{2\sigma^2}}\, dxdy$$

where $\sigma = 2\ x10^{-6}\ m$ , $W = 250\ x10^{-9}\ m$ and $L = 1\ x10^{-6}$ m . From the integral above, we obtain $\eta_{coupling} = 0.0098 \sim 1\ \%$. From literature [6], we take an absorption coefficient of 17 % for a BSCCO flake of 16.5 nm thickness. Within this rough estimation, the obtained detector efficiency for Device A at T = 20 K is $\eta = 3.3\ x\ 10^{-4}$ in the bias condition $I_b = 0.97\ I_C$. For $I_b = 0.85\ I_C$, the same analysis leads to $\eta = 3\ x\ 10^{-5}$. For Device B, at T = 15 K and $I_b = 0.98\ I_C$, the obtained detector efficiency is $\eta = 7.35\ x\ 10^{-5}$. Note that the extracted detection efficiency in the multiphoton regime does not account for the absorption probability for 2, 3 photons. Provided that multiple photons are absorbed in the nanodetector in a short time-period, the detector efficiency is comparable to the efficiency in the high-bias condition.

At this point, we briefly comment on possible strategies to improve the overall detector efficiency. Given the 2D nature of the active material, optimization of light-matter interaction in these devices should be a major point of emphasis in future developments. Optical cavities and nano-antennas can enhance the absorption probability in the device [7]. Detector designs with a higher fill factor, like Device C, could also improve the overall detector efficiency. However, the design of complex, high fill factor structures in ultrathin cuprates remains challenging. Integration into planar photonic architectures is also promising, as the coupling through evanescent modes might prove more efficient [2, 8].